# Decomposing GR(1) Games with Singleton Liveness Guarantees for Efficient Synthesis

Sumanth Dathathri          Richard M. Murray

*Abstract*— Temporal logic based synthesis approaches are often used to find trajectories that are correct-by-construction in systems – eg. synchronization for multi-agent hybrid systems, reactive motion planning for robots. However, the scalability of such approaches is of concern and at times a bottleneck when transitioning from theory to practice. In this paper, we identify a class of problems in the GR(1) fragment of linear-time temporal logic (LTL) where the synthesis problem allows for a decomposition that enables easy parallelization. This decomposition also reduces the alternation depth, resulting in more efficient synthesis. A multi-agent robot gridworld example with coordination tasks is presented to demonstrate the application of the developed ideas and also to perform empirical analysis for benchmarking the decomposition-based synthesis approach.

## I. INTRODUCTION

Robot motion planning has traditionally focused on generating trajectories from a given initial state to a final goal position. Recently, there has been increased attention towards generating trajectories with correct behavior in terms of synchronization of processes, safety and scheduling. The required behavior is often expressed in terms of a logic specification. The case of reactive robot motion planning from logic specifications involves considering complex behaviors for an adversarial environment and reasoning about all admissible behaviors for the environment to generate a plan for the robot. Temporal logic is often employed in this setting for reactive motion planning when we wish to generate motion plans that may exhibit complex behavior but are provably correct.

In particular, we focus on LTL [17], [19], a modal temporal logic often used as the mathematical language to formally specify the desired behavior for the robot [4], [10]. Synthesizing finite-memory strategies from LTL specifications for the general case is doubly exponential in the length of the formula [20], but for *Generalized Reactivity (1)* (GR (1))–a rich, expressive fragment of LTL, the synthesis can be done in polynomial time in the number of states and the number of liveness guarantees for the system and the number of liveness assumptions for the adversary [13]. GR(1) specifications model a game where the system and its adversary infinitely often satisfy a set of liveness constraints while making moves that satisfy certain safety constraints. This fragment in particular has received considerable attention since its conception because of the computational tractability associated with it. The GR(1) fragment is also particularly attractive because of the symbolic nature of the synthesis algorithm, that enables scaling to large finite-transition systems.

The complexity of synthesis for this class of temporal logic scales as cubic or quadratic [5] depending on the algorithm used for computing the fixed point. In our work, we identify a special subclass of GR(1) where we can provably decompose the synthesis problem into several smaller independent synthesis problems. During the decomposition procedure, we also eliminate one of the nested fixed points to reduce the *alternation depth*, thereby improving performance. In the context of motion planning, this class corresponds to finding paths that visit a set of nodes in a graph infinitely often, where each of the nodes in the graph satisfies a separate liveness condition. When an adversary is present, the relevant reactive motion planning problem is that of performing coordinated tasks for multi-agent systems [7]. Informally, the motion planning problem is to find a path for the robot such that when the environment is in pose $\mathcal{X}$, the robot has to be in pose $\mathcal{Y}$ where $(\mathcal{X}, \mathcal{Y})$ completes the coordination task. Figure 1 depicts such a problem instance where the controlled robot has to find a plan for its motion such that when the uncontrolled agent is at pose $E$, the robot has to attain pose $C$. When the environment is at pose $D$ (pose $B$), the robot must be at pose $C$ (pose $F$). For all assumed behaviors of the uncontrolled agent, the robot must (if feasible) find a reactive trajectory such that the agent-robot system visits the poses $(E, C)$, $(D, C)$ and $(B, F)$ infinitely often. For example, in [12], the authors propose an intermittent communication framework for mobile robot networks where the intermittent communication requirement is captured through an LTL formula that enforces the robots to meet infinitely often at certain rendezvous points.

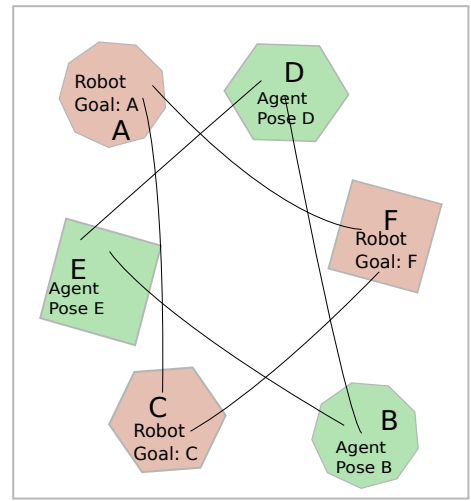

Fig. 1: Robot motion planning

The issue of scalability for synthesis algorithms has received attention in the past. In [2], the specification is compositionally decomposed to allow for scalable synthesis, with refinements being made to the decomposed specifications based on counter-strategies and strategies for the system and the environment. In [16], the GR(1) synthesis problem is decomposed into checking a feasibility problem and an online short-horizon strategy generation problem to improve scalability. In [21], the authors improve the scalability of controller synthesis by restricting themselves to linear systems and safety constraints, thereby enabling a reduction of the problem to the computation of control invariant sets. The main contribution of our work is to identify a subclass of GR(1) synthesis problems where the problem can be decomposed into smaller *reachability games*. The decomposed subgames are independent, allowing for the parallelization of the synthesis process.

## II. Preliminaries

In this section, we introduce game structures and $\mu$-calculus ([14]) over game structures as in [5] before presenting our contribution.

### A. Game Structures and $\mu$-calculus

A *game structure* $G = \langle \text{AP}, \text{AP}_{\text{env}}, \text{AP}_{\text{sys}}\, \theta^{\text{env}}, \theta^{\text{sys}}, \rho^{\text{sys}}, \rho^{\text{env}}, \varphi \rangle$ represents a two player game where

- $\text{AP} := \{v_1, \ldots, v_n\}$: A finite set of atomic propositions,
- $\text{AP}_{\text{env}} \subseteq \text{AP}$: Set of propositions controlled by the environment,
- $\text{AP}_{\text{sys}} \subseteq \text{AP}$: Set of atomic propositions controlled by the system. Also, $\text{AP}_{\text{env}} \cap \text{AP}_{\text{sys}} = \emptyset$.
- $\theta^{\text{sys}}, \theta^{\text{env}}$: Boolean formulas in $\text{AP}_{\text{sys}}$ and $\text{AP}_{\text{env}}$ characterizing the system and environment initial states,
- $\Sigma := 2^{\text{AP}}$,
- $\rho^{\text{env}}$ is a formula in the propositions $\text{AP}, \text{AP}_{\text{env}}$. An input $x \subseteq \text{AP}_{\text{env}}$ is a valid input at a state $s \in \Sigma$ if $(s, x) \models \rho^{\text{env}}$,
- $\rho^{\text{sys}}$ is a formula in the propositions $\text{AP}, \text{AP}_{\text{env}}, \text{AP}_{\text{sys}}$ that specifies the system transition rules. A system output/action $y \subseteq \text{AP}_{\text{sys}}$ is valid if $(s, x, y) \models \rho^{\text{sys}}$.

We will use the following Boolean operators: $\wedge$(conjunction), $\vee$(disjunction), $\rightarrow$(implication) and $\leftrightarrow$(bi-implication) to construct Boolean formulas. The temporal operators we use are *next* ($\bigcirc$), *eventually* ($\Diamond$) and *always* ($\square$).

In general, the semantics of LTL are defined over infinite words in $\Sigma^\omega$. However, to simplify notation, we extend the semantics of LTL to reason over finite strings. For a finite string $\gamma$ in $\Sigma^*$, we define:

$$\gamma \models \rho \Leftrightarrow \gamma\alpha \models \rho \text{ for any } \alpha \in \Sigma^\omega. \tag{1}$$

This allows for reasoning about states that might deadlock and playing games whose winning conditions allow for finite strings. For example, consider the formula $\Diamond \rho$. In accordance with the semantics of LTL, the language of this formula consists of strings in $\Sigma^\omega$ for which $\sigma_k \models \rho$ for some finite $k$. But, if we wanted to express the behavior for a motion planning problem where we are interested in only reaching a goal and not the behavior beyond, the extended semantics allow for that. According to the extended semantics of LTL for finite strings, the language of the above formula consists of strings in $\Sigma^\omega \cup \Sigma^*$ such that for $\sigma$ in the language of the above formula, there exists a finite $k$ such that $\sigma_k \models \rho$.

For a Boolean formula $\xi$, denote by $[[\xi]] \subseteq \Sigma$ the set of states that satisfy $\xi$. Additionally, given a game-structure $G$ and a Boolean formula $\psi$, we define the relevant $\mu$-calculus operator $\Diamondblack$ as:

$$[[\Diamondblack \psi]] = \{s \in \Sigma | \forall x \in \mathcal{P}(\text{AP}_{\text{env}}), (s, x) \models \rho^{\text{env}} \rightarrow \\ \exists y \in \mathcal{P}(\text{AP}_{\text{sys}}).(s, x, y) \models \rho^{\text{sys}} \wedge (x, y) \models \psi\}.$$

where $\mathcal{P}(\text{AP}_{\text{env}})$ denotes the power set of $\text{AP}_{\text{env}}$. $[[\Diamondblack \psi]]$ characterises the states from which the system can force the next state to satisfy $\psi$ for any valid input. In this paper, we use the subscript notation, e.g., $\sigma_0 \sigma_1 \sigma_2 \cdots \sigma_n \in \Sigma^*$, noting that infinite strings can also be regarded as functions of the natural numbers $\mathbb{N}$ into $\Sigma$. For a finite string $\gamma$, by $\gamma_{:,r}$ we refer to the the string $\gamma_0 \gamma_1 \ldots \gamma_{r-1}$. By $\gamma_{-1}$ we refer to the last element of $\gamma$. In other words, $\gamma_{-1} = \gamma_{|\gamma|-1}$.

### B. Definitions

Let $M$ be a finite set of memory values with $m^i \in M$ set to mark the initial memory value. A partial function $f: M \times \Sigma \times \mathcal{P}(\text{AP}_{\text{env}}) \rightarrow M \times \mathcal{P}(\text{AP}_{\text{sys}})$ is a finite-memory *strategy* for the game $G$ if for all $(w, s, x)$ for which $f(w, s, x)$ is defined, the condition $(s, x) \models \rho^{\text{env}} \rightarrow (s, x, y) \models \rho^{\text{sys}}$ holds.

A *play* $\sigma \in \Sigma^\omega \cup \Sigma^*$ for a strategy $f$ is the maximimal sequence of states such that $\exists m \in M^\omega$ with $m_0 = m^i$ such that $(m_{k+1}, \sigma_{k+1} \cap \text{AP}_{\text{sys}}) = f(m_k, \sigma_k, \sigma_{k+1} \cap \text{AP}_{\text{env}})$ and $\sigma_k \sigma_{k+1} \models \rho^{\text{env}}$ for $|\sigma| \geq k - 1 \geq 0$ if $\sigma \in \Sigma^*$ and $k \geq 0$ if $\sigma \in \Sigma^\omega$. By a sequence being maximal we imply that the play terminates when we reach a state where the strategy is not defined for a valid input or the environment deadlocks i.e. there is no valid input. We denote by $\text{Plays}(f)$ the set of all plays generated by $f$. Also, define the set $\text{Pref}(f) \in \Sigma^*$ as:

$$\text{Pref}(f) := \{\sigma \in \Sigma^* | \exists \sigma\gamma \in \text{Plays}(f).\gamma \in \Sigma^\omega \cup \Sigma^*\}.$$

Denote by $m^{\sigma, f}$ the sequence of memory values generated by $f$ corresponding to $\sigma \in \text{Plays}(f) \cup \text{Pref}(f)$. That is, we start with the initial memory value and update the memory values as indicated by $f$. More formally,

$$(m^{\sigma, f}_{k+1}, \sigma_{k+1} \cap \text{AP}_{\text{sys}}) = f(m^{\sigma, f}_k, \sigma_k, \sigma_{k+1} \cap \text{AP}_{\text{env}})$$

Given a strategy $f: M \times \Sigma \times \mathcal{P}(\text{AP}_{\text{env}}) \rightarrow M \times \mathcal{P}(\text{AP}_{\text{sys}})$, we define the set of reachable state memory pairs from an initial condition $\theta = \theta^{\text{sys}} \wedge \theta^{\text{env}}$ and an initial memory value $m^i$:

$$\{(w, s) | \exists j : (w, s) = (m^{\sigma, f}_j, \sigma_j) : \sigma \in \text{Plays}(f), \\ \sigma_0 \models \theta, m^{\sigma, f}_0 = m^i\}.$$

For a state $s$ in $\Sigma$, a strategy $f$ is *winning* if

$$\forall \sigma \in \text{Plays}(f).(\sigma_0 = s \to \sigma \models \varphi), \quad (2)$$

$f(m_{-1}^{\sigma,f}, \sigma_{-1}, x)$ is defined $\forall \sigma \in \text{Pref}(f)$ with $\sigma_0 = s$,

$$\forall x \in \text{AP}_{\text{env}} \text{ with } \sigma_{-1}x \models \rho^{\text{env}}. \quad (3)$$

with $\varphi$ being the winning condition. Note that from the above definition the system has a well-defined output action at every state-memory pair visited by following $f$, for all valid inputs, starting from a winning state, as long as the environment has not violated the safety assumption in the past. Dually, we define a winning state based on the existence of a winning strategy. A state $s \in \Sigma$ is a winning state against a condition $\varphi$ if there exists a strategy $f$ that is winning from that state. The winning set is the largest set of winning states from which there exists a strategy $f$ that is winning. We use $W_\varphi$ to denote the set of winning states associated with a formula $\varphi$.

### C. Generalized Reactivity(1)

A game structure G with a winning condition of the form

$$\varphi := \bigwedge_{i=1}^{m} \square \lozenge \psi_i^{\text{env}} \to \bigwedge_{j=1}^{n} \square \lozenge \psi_j^{\text{sys}} \quad (4)$$

is an instance of a *Generalized Reactivity(1)* game where $\psi_i^{\text{env}}$ and $\psi_j^{\text{sys}}$ are Boolean formula in AP.

*Problem 1:* For a given game structure $G$, the *GR(1) synthesis* problem is to find a finite memory strategy $f$ that is winning for the set of states satisfying the initial condition $\theta$, against the condition $\varphi$ (as in equation (4)).

The worst case complexity for GR(1) synthesis in general scales as $\mathcal{O}\left((nm|\Sigma|)^3\right)$ [13]. Using the approach in [6], a GR(1) game can be solved with $\mathcal{O}(nm|\Sigma|^2)$ *next step* computations but this memoization scheme stores many binary decision diagrams (BDDs) simultaneously which makes reordering the BDDs expensive. We refer the reader to [3] for an introduction to BDDs, and to [1], [18] for a discussion of variable reordering algorithms.

For the case when all the liveness formulae $\psi_i^{\text{sys}}$ are such that the sets $[[\psi_i^{\text{sys}}]]$ are singletons i.e there exists exactly one $s \in \Sigma$ for each $\psi_i^{\text{sys}}$ such that $s \models \psi_i^{\text{sys}}$, we propose an approach to decompose the original GR(1) games into $n + 1$ independent smaller subgames. Solving each smaller subgame involves solving a $\mu$-calculus formula with a smaller alternation depth of 2. The *alternation depth* of a formula is the number of alternations in the nesting of least and greatest fixpoints.

## III. DECOMPOSITION FOR SINGLETON LIVENESS GUARANTEES

### A. Reachability Games

A *reachability* game is an instance of a game structure G with a winning condition of the form

$$\varphi_{\text{rg}} := \bigwedge_{i=1}^{m} \square \lozenge \psi_i^{\text{env}} \to \lozenge \psi^{\text{sys}}. \quad (5)$$

$L(\varphi_{rg}) = \{\sigma : \sigma \in \Sigma^* \cup \Sigma^\omega, \exists \text{ finite } k \text{ such that } \sigma_k \models \psi^{\text{sys}} \text{ OR } \sigma \in \Sigma^\omega \text{ such that } \sigma \models \bigvee_{i=1}^{m} \lozenge \square \neg \psi_i^{\text{env}}\}$.

*Remark 1:* For a game structure G, the winning states for a reachability game can be computed by solving a $\mu$-calculus formula with an alternation depth of 2.

This holds as a direct consequence of Lemma 9 from [13]. Consider the $\mu$-calculus formula $\mu_{\text{rg}}$ defined as:

$$\mu_{\text{rg}} := \mu Y \left( \bigvee_{j=1}^{m} \nu X \left( \left( (\psi^{\text{sys}} \vee \lozenge Y) \vee \neg \psi_j^{\text{env}} \right) \wedge \lozenge X \right) \right). \quad (6)$$

Here, $\nu$ is the greatest fixpoint operator and $\mu$ is the least fixpoint operator (See [22] for detailed definitions of these operators). The alternation depth for this formula is 2. Intuitively, the fixed point in $X$ characterizes the set of states from which the system can force the play to stay indefinitely in $[[\neg \psi_j^{\text{env}}]]$ for some $j$ or in a finite number of steps reach a state satisfying $\psi^{\text{sys}} \vee \lozenge Y$. Staying in $[[\neg \psi_j^{\text{env}}]]$ for some $j$ indefinitely implies blocking the environment from satisfying one of its liveness assumptions. The outer least fixed point in $Y$ makes sure that the phase of play represented by $\lozenge Y$ eventually ends in $[[\psi^{\text{sys}}]]$. This way either $\lozenge \psi^{\text{sys}}$ is satisfied or $\bigvee_{i=1}^{m} \lozenge \square \neg \psi_i^{\text{env}}$ is satisfied. These fixed points can be computed with complexity $\mathcal{O}\left((m|\Sigma|)^2\right)$[9]. The approach in [6] results in $\mathcal{O}(m|\Sigma|^2)$ *next step* computations to solve for the fixed points.

### B. Generalized Reactivity (1) Games

In this section, we identify a special class of $GR(1)$ synthesis problems where the winning strategy can be computed by solving $n + 1$ reachability games instead of solving the cyclic $\mu$-calculus formula with an alternation depth of 3.

For the case with two liveness guarantees, the $\mu$-calculus formula in [13] can be written using the *vector notation* as:

$$\mu_\varphi = \nu \begin{bmatrix} Z_1 \\ Z_2 \end{bmatrix} \begin{bmatrix} \mu Y \left( \bigvee_{j=1}^{m} \nu X \left( ((\psi_1^{\text{sys}} \wedge \lozenge Z_2) \vee \lozenge Y \vee \neg \psi_j^{\text{env}}) \wedge \lozenge X \right) \right) \\ \mu Y \left( \bigvee_{j=1}^{m} \nu X \left( ((\psi_2^{\text{sys}} \wedge \lozenge Z_1) \vee \lozenge Y \vee \neg \psi_j^{\text{env}}) \wedge \lozenge X \right) \right) \end{bmatrix} \quad (7)$$

The $\mu$-calculus formula in equation (7) has an alternating depth of 3. We propose and prove an algorithm to solve for the winning states and extract a strategy by solving independent reachability-games, each involving solving a $\mu$-calculus formula with an alternation depth of 2.

A GR(1) game with $n$ liveness guarantees is decomposed into $n + 1$ reachability games when $|[[\psi_i^{\text{sys}}]]| = 1$ for each $i$. The reachability games are independent, unlike the cyclic dependency in equation (7) between the various liveness guarantees. The outermost fixed point computation in $Z_i$ can be avoided here as the liveness guarantees correspond to singleton sets and this allows for the separation of the subgames (we prove this later). Here, for example, the GR(1)

game can be split into three reachability games:

$$\bigwedge_{i=1}^{m} \Box \Diamond \psi_i^{\text{env}} \to \Diamond \psi_1^{\text{sys}}, \quad (8)$$

$$\bigwedge_{i=1}^{m} \Box \Diamond \psi_i^{\text{env}} \to \Diamond \psi_2^{\text{sys}},$$

$$\bigwedge_{i=1}^{m} \Box \Diamond \psi_i^{\text{env}} \to \Diamond \texttt{False},$$

with the initial conditions $\theta_1 = \psi_2^{\text{sys}}$, $\theta_2 = \psi_1^{\text{sys}} \vee \theta$ and $\theta_0 = \theta$ from the reachability games respectively (recall $\theta$ is the initial condition for the original synthesis problem). In general, for a problem with $n$ liveness constraints, the reachability games can be set-up as for $j \in \{1, 2, \ldots, n\}$:

$$\bigwedge_{i=1}^{m} \Box \Diamond \psi_i^{\text{env}} \to \Diamond \psi_{j \oplus 1}^{\text{sys}} \quad (9)$$

with the initial conditions being $\theta_j = \psi_j^{\text{sys}}$ for $j \neq n$ and $\theta_j = \psi_j^{\text{sys}} \vee \theta$ for $j = n$. Note that $\oplus$ is the modulo $n$ operator i.e $j \oplus 1 = (j+1)$ modulo $n$. For example, $n \oplus 3 = 3$ when $3 < n$.

We shall refer to the winning condition

$$\bigwedge_{i=1}^{m} \Box \Diamond \psi_i^{\text{env}} \to \Diamond \psi_k^{\text{sys}}$$

as $\varphi_k^{\text{reach}}$. Additionally, define $\varphi_0^{\text{reach}}$ as

$$\bigwedge_{i=1}^{m} \Box \Diamond \psi_i^{\text{env}} \to \Diamond \texttt{False} \quad (10)$$

and the initial condition for this game is $\theta$. Note that for any given state, a strategy that is winning against this condition can only do so by forcing the play to block the environment from satisfying its assumptions.

## IV. STRATEGY FOR GR(1) FROM REACHABILITY GAMES

Here, we formalize the approach for combining the strategies for the reachability games. Suppose $\varphi_0^{\text{reach}}$ is winnable, then from solving $\varphi_0^{\text{reach}}$ we have a strategy for $\bar{\varphi}$ and this is also winning for $\varphi$, since the environment is blocked from satisfying its assumptions.

Suppose $\varphi_0^{\text{reach}}$ is not winnable and the other $n$ reachability games are winnable. We construct the strategy $f_G^{\varphi}$ by combining the $n$ reachability games that is winning against $\varphi$. To do this, we introduce a variable $\mathcal{Z}_n$ that can take values in $\{1, 2, \ldots, n\}$ to track which liveness guarantees have been satisfied in the current cycle, with $\mathcal{Z}_n$ initialized to $n$. Let $f_i^{reach}$ be the winning strategy for $\varphi_i^{\text{reach}}$. The strategy $f_G^{\varphi}$ is constructed such that starting with a state $s \models \theta$ the execution follows $f_n^{\text{reach}}$ to reach a state satisfying $\psi_1^{\text{sys}}$ or blocks the environment from satisfying one of the liveness assumptions. If the execution reaches $\psi_1^{\text{sys}}$, the strategy switches to $f_1^{\text{reach}}$ and reaches $\psi_2^{\text{sys}}$ or blocks the environment, and so on. Formally, the strategy

$$f_G^{\varphi} : (M \times \{1, 2, \ldots, n\}) \times \Sigma \times \mathcal{P}(\text{AP}_{\text{env}}) \to$$

$$(M \times \{1, 2, \ldots, n\}) \times \mathcal{P}(\text{AP}_{\text{sys}})$$

is constructed as

$$f_G^{\varphi}((w, \mathcal{Z}_n), s, s' \cap \text{AP}_{\text{env}}) = ((w', \mathcal{Z}_n'), s' \cap \text{AP}_{\text{sys}}),$$

where if $s \models \psi_{\mathcal{Z}_n \oplus 1}^{\text{sys}}$,

$$\mathcal{Z}_n' = \mathcal{Z}_n \oplus 1,$$

$$(w', s' \cap \text{AP}_{\text{sys}}) = f^{\text{reach}\mathcal{Z}_n'}(m_0^{\mathcal{Z}_n'}, s, s' \cap \text{AP}_{\text{env}},),$$

and if $s \not\models \psi_{\mathcal{Z}_n \oplus 1}^{\text{sys}}$,

$$(w', s' \cap \text{AP}_{\text{sys}}) = f^{\text{reach}\mathcal{Z}_n}(w, s, s' \cap \text{AP}_{\text{env}},),$$

$$\mathcal{Z}_n' = \mathcal{Z}_n.$$

Here $\mathcal{Z}_n'$ denotes the value of $\mathcal{Z}_n$ at the next step. Similarly, $s'$ is the the next state with $s$ being the current state. When $s \models \psi_{\mathcal{Z}_n \oplus 1}^{\text{sys}}$ for a given $\mathcal{Z}_n$, we increment $\mathcal{Z}_n$. Thereby switching to the strategy $f^{\text{reach}\mathcal{Z}_n \oplus 1}$ till we reach $\psi_{\mathcal{Z}_n \oplus 2}^{\text{sys}}$.

If for the initial condition $\theta$, $\varphi_0^{\text{reach}}$ is not winnable and for some $i$ such that $n \geq i > 0$, $\varphi_i^{\text{reach}}$ is not winnable then $\varphi$ is not winnable from $\theta$.

## V. DISCUSSION

In general GR(1) games do not allow for such a decomposition, but the singleton nature of the sets corresponding to the liveness goals allows us to decompose the specifications here. To see why, we take a closer look at the $\mu$-calculus formula correspoding to a GR(1) game with two liveness guarantees for the system from equation (7).

To simplify things, suppose that the environment cannot be blocked from satisfying its liveness assumptions (i.e. $\varphi_0^{\text{reach}}$ is not reazliable). The fixed points in $Z_1$ and $Z_2$ are such that after starting at a state in $Z_1$, the system forces the play into $Z_2$ while visiting a state satisfying $\psi_1^{\text{sys}}$. Similarly, from $Z_2$ the system forces the play into a state in $Z_1$ while visiting $\psi_2^{\text{sys}}$. This could take a few iterations for the fixed points in $Z_1$ and $Z_2$ to converge because a change in $Z_1$ may cause a change in $Z_2$, and this goes on till the greatest fixed point is met. Figure 2 illustrates this process, where initially $Z_2$ is the set of all states and $Z_1$ is the set of states from which the system can force the play into $Z_2$ from visiting Goal 1. Subsequently, in the next computation, $Z_2$ is the set of states from where the play can be forced into $Z_1$ after visitng Goal 2. The iterations are continued till convergence [1].

But, if the sets corresponding to the liveness guarantees are singletons, we can chain a sequence of strategies (synthesized independently) at the liveness guarantees to enforce the cyclic satisfaction of the liveness guarantees as opposed to going through the iterations in $Z_i$.

---

[1]In the case when the liveness guarantees correspond to singleton sets, it can be shown that no more than two iterations are needed to compute the outermost fixpoint.

[2]$Z_2$ after iteration $i$ is a subset of $Z_1$ from iteration $i$, $Z_1$ from iteration $i + 1$ is a subset of $Z_1$ from iteration $i$. The figure is not an accurate representation and is drawn so for better visualization.

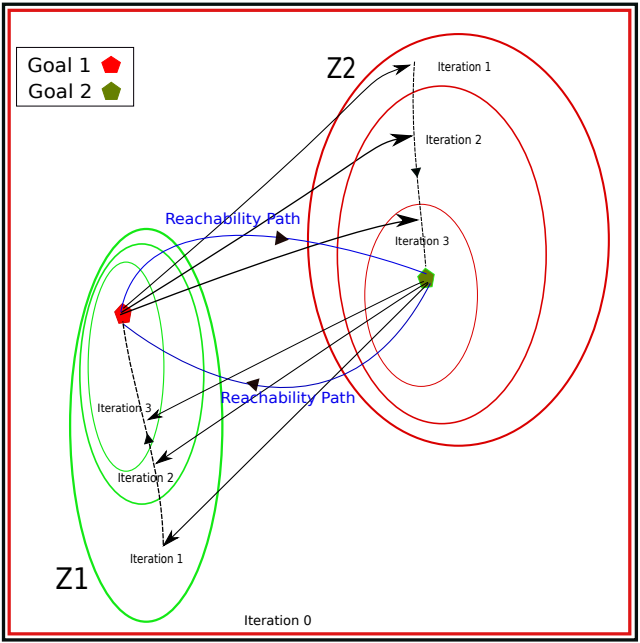

Fig. 2: Visualization of the different approaches [2]

## VI. RESULTS

Define $\bar{\varphi}$ as the following formula:

$$\bar{\varphi} := \bigwedge_{i=1}^{m} \psi_i^{\text{env}} \to \Diamond \psi_1^{\text{sys}} \wedge \left( \bigwedge_{i=1}^{n} \Diamond \left( \psi_i^{\text{sys}} \to \Diamond \psi_{i \oplus 1}^{\text{sys}} \right) \right). \quad (11)$$

*Claim 2:* $W_\varphi = W_{\bar{\varphi}}$ if $|[[\psi_i^{\text{sys}}]]| = 1 \, \forall i \in \{1, 2 \ldots n\}$.
The winning sets for the formulas $\bar{\varphi}$ and $\varphi$ are the same. This implies that the set of states from which we can satisfy each $\psi_i^{\text{sys}}$ once is the same as the set of states from which we can cycle through the $\psi_i^{\text{sys}}$-s infinitely often. A proof of the Claim is provided in Appendix I.

*Lemma 3:* A game structure G with a GR(1) winning condition of the form $\varphi = \bigwedge_{i=1}^{m} \Box \Diamond \psi_i^{\text{env}} \to \bigwedge_{i=1}^{n} \Box \Diamond \psi_i^{\text{sys}}$ can be solved by solving $n + 1$ independent reachability games if $|[[\psi_i^{\text{sys}}]]| = 1 \, \forall i \in \{1, 2, \ldots, n\}$.
A proof of the Lemma is provided in Appendix II.

Note that the construction of the strategy in Section IV combines constructions from the proofs of Claim 2 and Lemma 3.

## VII. EXPERIMENTS

For comparing the performance of the synthesis algorithm proposed here with standard GR(1) synthesis, we consider the problem of coordinated planar reactive robot motion planning on a gridworld. For a given set of cells $\{(a_1^r, a_1^c), (a_2^r, a_2^c), \ldots, (a_n^r, a_n^c)\}$ and $\{(b_1^r, b_1^c), (b_2^r, b_2^c), \ldots, (b_n^r, b_n^c)\}$, the controlled robot has to coordinate with a moving agent such that the robot is in cell $(b_i^r, b_i^c)$ when the agent is in cell $(a_i^r, a_i^c)$. The robot has to complete this coordination task infinitely often. To make sure the problem is feasible, the cells $\{\{(b_1^r, b_1^c), \ldots, (b_n^r, b_n^c)\}\}$ are added as liveness conditions for the agent. The robot's motion constraints allow movement to any of its non-diagonally adjacent cells.

Let $Y_r$ denote the row (horizontal) position of the controlled robot $Y_c$ the column (vertical) position. Similarly, let $X_r$ and $X_c$ denote the row and the column position of the uncontrolled agent. The transition rule for the robot at position $(Y_r, Y_c) = (i, j)$ can be written as:

$$(Y_r = i \wedge Y_c = j) \to \Big( (Y_r' = i + 1 \wedge Y_c' = j) \\ \vee (Y_r' = i \wedge Y_c' = j + 1) \\ \vee (Y_r' = i \wedge Y_c' = j) \\ \vee (Y_r' = i - 1 \wedge Y_c' = j) \\ \vee (Y_r' = i \wedge Y_c' = j - 1) \Big)$$

with the additional constraint that $Y_r$ and $Y_c$ always stay in the bounds of the gridworld i.e

$$0 \leq Y_r \leq r_{\max}, 0 \leq Y_c \leq c_{\max}.$$

The agent's motion is constrained in a similar way. Furthermore, the controlled robot as a part of the safety specification has to avoid collision with the uncontrolled agent i.e

$$\neg(X_r = Y_r \wedge Y_c = X_c)$$

where $\neg$ is the negation operator. Both the controlled robot and the uncontrolled agent have to avoid collisions the walls (shaded). For example, if the location $(w_r, w_c)$ is shaded, then the safety specification corresponding to avoiding collision with this wall for the controlled agent is:

$$\neg(Y_r = w_r \wedge Y_c = w_c).$$

The liveness assumptions can be specified as:

$$\bigwedge_{i=1}^{n} \Box \Diamond (X_r = b_i^r \wedge X_c = b_i^c).$$

The liveness guarantees are written as:

$$\bigwedge_{i=1}^{n} \Box \Diamond (Y_r = a_i^r \wedge Y_c = a_i^c \wedge X_r = b_i^r \wedge X_c = b_i^c).$$

Figure 3 shows an example gridworld instance. The runtimes for the approach presented here using the decomposed reachability games is compared with those for the solvers gr1c[15] and slugs[8]. The solvers are accessed using the interfaces in the Temporal Logic Planning Toolbox (TuLiP)[11]. The reachability games are solved using the rg module from gr1c. The computations were performed on a 2.40GHz Quadcore machine with 16 GB of RAM [3].

Gridworld instances with varying number of liveness guarantees and gridsizes are used for benchmarking. Figure 4 depicts the mean runtimes from the benchmarking experiments on $t \times t$– sized gridworld instances, with varying $t$. For each grid size, 50 random problem instances (with a wall density of 10 percent and 6 liveness guarantees) are created. We see that the decomposition based approach outperforms GR(1) synthesis (using `slugs` and `gr1c`).

---

[3]Code for the implemented examples is available at https://www.dropbox.com/sh/9pfqysj2z572f5b/AAATa7VlGmpT9ljpCRgyIwhRa?dl=0

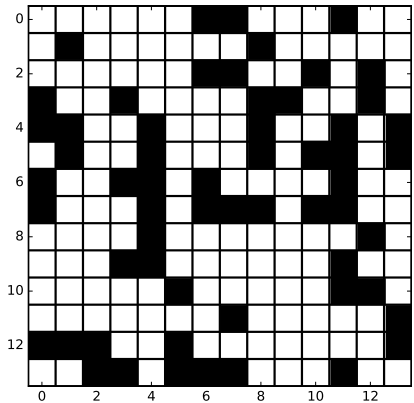

Fig. 3: Gridworld of size $14 \times 14$ with wall density of 0.3

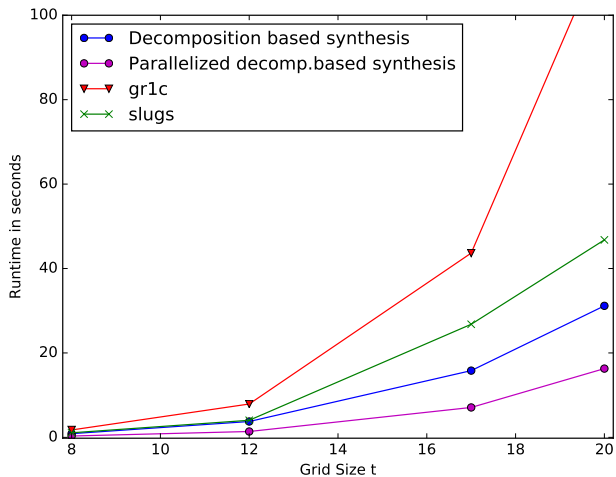

Fig. 4: Performance on gridworld problems with varying grid size ($t \times t$)

Figure 5 depicts the performance for gridworld instances of size $14 \times 14$ (wall density 0.3) with the number of liveness constraints changing. Here again we observe similar trends with the decomposition approach outperforming GR(1) synthesis using `slugs` and `gr1c`. When the reachability games are solved in parallel, we observe improved scaling for the decomposition based approach. The slope for the parallelized decompositioned-based synthesis is lesser than that of decomposition-based synthesis without parallelization. This is because the complexity of each of the reachability games is independent of $n$, where $n$ is the number of liveness-guarantees. Since the $n+1$ reachability games are solved in parallel, the runtime approximately stays constant even with the varying number of liveness guarantees.

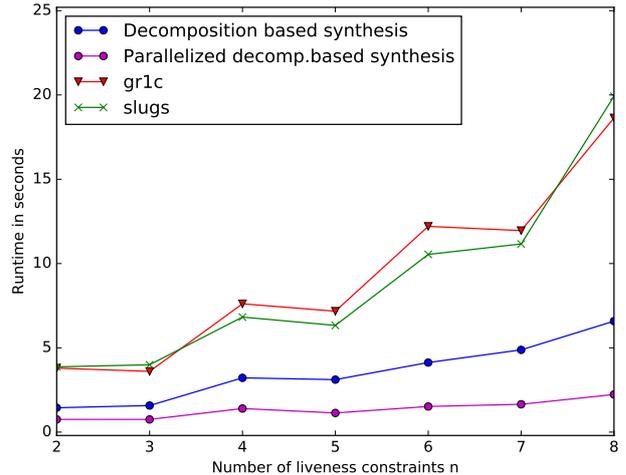

Fig. 5: Performance on gridworld problems with varying number of liveness constraints

## VIII. CONCLUSION

We identified a class of synthesis problems where GR(1) games can be solved by solving a set of reachability games that arise from decomposing the original GR(1) formula. The decomposed reachability games are independent from each other allowing for parallelization. Synthesis from the decomposed set of games also results in better performance from eliminating the outer greatest fixed point in GR(1) synthesis. For benchmarking, the decomposition based approach is used to solve LTL motion planning problems that involve coordination between an external agent and a controlled robot on a gridworld. The experiments demonstrate the improved performance of the decomposition-based approach.

Directions for future work include extending the results here to the case when one of the liveness guarantees corresponds to a set of states that is a singleton. Even though in that case the problem cannot be fully decomposed for parallelization–the alternation depth can still be reduced by eliminating the outer greatest fixed point corresponding to GR(1) synthesis. Using the intuiton from here, we wish to further explore approaches for the decomposition of the GR(1) synthesis problems in more general settings to allow for parallelization–heuristic based if complete decomposition is not possible. Another direction of research is to explore approaches for abstraction where

1) Existing discrete transition systems can be abstracted into those with liveness conditions that correspond to singleton sets;
2) For abstraction based synthesis for systems with continous dynamics, the abstraction into a discrete transition system is done so that the sets corresponding to the liveness guarantees are singletons.

*Acknowledgments:* The authors would like to thank Scott C. Livingston and Tung M. Phan for helpful input. This work was supported by STARnet, a Semiconductor Research Corporation program, sponsored by MARCO and DARPA.

## Appendix I
## Proof of Claim 2

*Proof:* First we show $W_\varphi \subseteq W_{\bar\varphi}$. Let $f_G^\varphi$ be a winning strategy for the condition $\varphi$ for the set $W_\varphi$. We show that $f_G^\varphi$ is winning for the condition $\bar\varphi$ for the set $W_\varphi$, thereby proving that $W_\varphi \subseteq W_{\bar\varphi}$.

Consider $\sigma \in \text{Plays}(f_G^\varphi)$ such that $\sigma_0 \in W_\varphi$. By definition,
$$\sigma \models \left( \bigwedge_{i=1}^m \Box\Diamond \psi_i^{\text{env}} \to \bigwedge_{j=1}^n \Box\Diamond \psi_j^{\text{sys}} \right).$$

If $\sigma \models \bigvee_{i=1}^m \Diamond\Box\neg\psi_i^{\text{env}}$, then $\sigma \models \bar\varphi$ directly. In the other case,
$$\sigma \models \left( \bigwedge_{j=1}^n \Box\Diamond \psi_j^{\text{sys}} \right)$$
has to hold. For this case, the semantics of the $\Diamond$ operator imply that there exists $k_1, k_2, \ldots, k_n$ that are finite such that $\sigma_{k_i} \models \psi_i^{\text{sys}}$ ($\Box\Diamond\psi_i^{\text{sys}}$ implies $\Diamond\psi_i^{\text{sys}}$ has to hold at every step along a run). For each such finite $k_i$, $\forall j \in \{1, 2, \ldots, n\} \exists h_{ij} > k_i . \sigma_{h_{ij}} \models \psi_j^{\text{sys}}$. This implies $\Diamond(\psi_i^{\text{sys}} \wedge \Diamond\psi_{i\oplus 1}^{\text{sys}})$ holds for each $i$. Therefore, $\sigma_0 \in W_{\bar\varphi}$ and hence $W_\varphi \subseteq W_{\bar\varphi}$.

Now, let us show $W_{\bar\varphi} \subseteq W_\varphi$. $W_{\bar\varphi}$ is the winning set for $\bar\varphi$. By definition of winning sets there exists a winning strategy $f_G^{\bar\varphi}$ that is winning against $\bar\varphi$ for every element of $W_{\bar\varphi}$. Also, $W_{\bar\varphi}$ is not an empty set if the system can win for $\bar\varphi$ from any state. If $W_{\bar\varphi}$ is an empty set, $W_{\bar\varphi} \subseteq W_\varphi$ is trivially true.

To prove that $W_{\bar\varphi} \subseteq W_\varphi$, we construct a new strategy $\bar f$ that is winning against the condition $\varphi$ for all states in $W_{\bar\varphi}$. This way we show that every state in $W_{\bar\varphi}$ is winning against $\varphi$ and hence in $W_\varphi$.

Consider some play $\bar\sigma$ of $f_G^{\bar\varphi}$ such that $\bar\sigma \models \bigwedge_{i=1}^m \Box\Diamond\psi_i^{\text{env}}$ and $\bar\sigma_0 \in W_{\bar\varphi}$. If no such play exists, then for all plays of $f_G^{\bar\varphi}$, the condition $\neg \bigwedge_{i=1}^m \Box\Diamond\psi_i^{\text{env}}$ holds and the strategy $f_G^{\bar\varphi}$ is winning for $\varphi$ because all plays of $f_G^{\bar\varphi}$ satisfy $\varphi$.

Consider the case when such a play exists. $\Diamond\psi_1^{\text{sys}}$ holds implying that at some finite $k$, $\bar\sigma_{k:} \models \psi_1^{\text{sys}} \wedge \Diamond\psi_{1\oplus 1}^{\text{sys}}$ holds. Denote the smallest $k$ at which $\bar\sigma_k \models \psi_1^{\text{sys}} \wedge \Diamond\psi_{1\oplus 1}^{\text{sys}}$ as $k_1$. By a similar reasoning, we can go on to define $k_1, k_2, \ldots, k_n$. Next, we introduce a variable $\mathcal{Z}_n$ that can take values in $\{1, 2, \ldots, n\}$ and tracks which of the liveness guarantees have been satisfied. $\mathcal{Z}_n$ is initialized to 1. The strategy
$$\bar f : (M \times \{1, 2, \ldots, n\}) \times \Sigma \times \mathcal{P}(\text{AP}_{\text{env}}) \to$$
$$(M \times \{1, 2, \ldots, n\}) \times \mathcal{P}(\text{AP}_{\text{sys}})$$
is constructed as
$$\bar f((w, \mathcal{Z}_n), s, s' \cap \text{AP}_{\text{env}}) = ((w', \mathcal{Z}_n'), s' \cap \text{AP}_{\text{sys}}),$$
where if $s \models \psi_{\mathcal{Z}_n}^{\text{sys}}$,
$$(w', s' \cap \text{AP}_{\text{sys}}) = f_G^{\bar\varphi}(m_{k_{\mathcal{Z}_n}}^{\bar\sigma, f_G^{\bar\varphi}}, s, s' \cap \text{AP}_{\text{env}}),$$
$$\mathcal{Z}_n' = \mathcal{Z}_n \oplus 1,$$
and if $s \not\models \psi_{\mathcal{Z}_n}^{\text{sys}}$,
$$(w', s' \cap \text{AP}_{\text{sys}}) = f_G^{\bar\varphi}(w, s, s' \cap \text{AP}_{\text{env}}),$$
$$\mathcal{Z}_n' = \mathcal{Z}_n.$$

*a) Showing well-definedness for all relevant inputs:* For any reachable state-memory pair $(s,w)$ of $f_G^{\bar{\varphi}}$ and any input $x \in \mathcal{P}(\mathrm{AP}_{\mathrm{env}})$, $f_G^{\bar{\varphi}}(w,s,x)$ is defined if $(s,x) \models \rho^{\mathrm{env}}$ (since $f_G^{\bar{\varphi}}$ is winning for $\bar{\varphi}$). For the case when $(s,w)$ is reachable, then $\bar{f}(w,s,x)$ is also reachable if $sx \models \rho^{\mathrm{env}}$. This implies that when $s \not\models \psi_{\mathcal{Z}_n}^{\mathrm{sys}}$ if $ss' \models \rho^{\mathrm{env}}$ and $(s,w)$ is reachable, then $(w', s')$ with $(w', s' \cap \mathrm{AP}_{\mathrm{sys}}) = \bar{f}(w, s, s' \cap \mathrm{AP}_{\mathrm{env}})$ is reachable.

Consider a state $s$ such that $s \models \psi_{\mathcal{Z}_n}^{\mathrm{sys}}$, $s = \bar{\sigma}_{k_{\mathcal{Z}_n}}$ because $\bar{\sigma}_{k_{\mathcal{Z}_n}} \models \psi_{\mathcal{Z}_n}^{\mathrm{sys}}$ and $[[\psi_{\mathcal{Z}_n}^{\mathrm{sys}}]]$ is a singleton. If $s = \bar{\sigma}_{k_{\mathcal{Z}_n}}$, then $f_G^{\bar{\varphi}}(m_{k_{\mathcal{Z}_n}}^{\bar{\sigma}, f_G^{\bar{\varphi}}}, s, x)$ is defined $\forall x \in \mathcal{P}(\mathrm{AP}_{\mathrm{env}}).sx \models \rho^{\mathrm{env}}$. This is because $(s, m_{k_{\mathcal{Z}_n}}^{\bar{\sigma}, f_G^{\bar{\varphi}}})$ is reached during the execution $\bar{\sigma} \in Plays(f_G^{\bar{\varphi}})$. Therefore, for any $s \models \psi_{\mathcal{Z}_n}^{\mathrm{sys}}$, $\bar{f}(m_{k_{\mathcal{Z}_n}}^{\bar{\sigma}, f_G^{\bar{\varphi}}}, s, (.))$ is well-defined for all valid environmental inputs and $(s, m_{k_{\mathcal{Z}_n}}^{\bar{\sigma}, f_G^{\bar{\varphi}}})$ is reachable for $f_G^{\bar{\varphi}}$.

Additionally, we begin execution for the first input at an initial memory value $m^i \in M$. For a valid initial state $s \in W_{\bar{\varphi}}$ and the initial memory value $m^i$, $(s, m^i)$ is reachable for $f_G^{\bar{\varphi}}$. To summarize, we start at a reachable state-memory pair for $f_G^{\bar{\varphi}}$.

We showed that for any reachable state-memory pair $(s, w)$ of $f_G^{\bar{\varphi}}$, $\bar{f}$ is well-defined for all valid environmental inputs. We also showed that the output for this case is a reachable state-memory pair (for $f_G^{\bar{\varphi}}$) if the environmental input is valid. Additionally, we also start at a reachable state-memory pair. Therefore, for any $\sigma \in \mathrm{Pref}(\bar{f})$, at $(\sigma_{-1}, m_{-1}^{\sigma, f})$, $\bar{f}$ is well-defined for all valid inputs if $\sigma_r \sigma_{r+1} \models \rho^{\mathrm{env}}$ $\forall r < |\sigma| - 1$, $\sigma_0 \in W_{\bar{\varphi}}$, and execution starts with the initial memory value $m^i$.

*b) Proving properties about the strategy $\bar{f}$:* We argued that $\bar{f}$ satisfies the condition in equation (3). This implies that for a state, $\bar{f}$ is well-defined for any valid environmental input when the environment assumption has not been violated in the past while getting to that state. Now all that remains is to show that the plays of $\bar{f}$ satisfy the specification $\varphi$.

Consider any $\sigma \in \mathrm{Plays}(\bar{f})$ with $\sigma_0 \in W_{\bar{\varphi}}$. Note that the state sequence $\bar{\sigma}$ used for the construction of the strategy $\bar{f}_G^{\varphi}$ is independent of the sequence of inputs corresponding to $\sigma$. Also, note that the strategy $\bar{f}$ and $\mathrm{Plays}(\bar{f})$ have already been defined. Here we only prove properties about elements of the set $\mathrm{Plays}(\bar{f})$, specifically that they satisfy $\varphi$. Consider the case when $\sigma \models \bigwedge_{i=1}^{m} \Box \Diamond \psi_i^{\mathrm{env}}$, because for the other case $\varphi$ holds directly.

Execution begins at a valid initial state $m^i$ and $\sigma_0 \in W_{\bar{\varphi}}$. If $\sigma \not\models \Diamond \psi_1^{\mathrm{sys}}$, it implies that execution continued in accordance with $f_G^{\bar{\varphi}}$ without any memory resets (from the definition of $\bar{f}$). This implies that $\sigma \in \mathrm{Plays}(f_G^{\bar{\varphi}})$, but this leads to a contradiction since $\sigma \models \Diamond \psi_1^{\mathrm{sys}} \wedge \bigwedge_{i=1}^{n} \Diamond(\psi_i^{\mathrm{sys}} \to \Diamond \psi_{i\oplus 1}^{\mathrm{sys}})$, implying $\sigma \models \Diamond \psi_1^{\mathrm{sys}}$. This is because $\sigma \models \bar{\varphi}$ and we are looking at the case when $\sigma \models \bigwedge_{i=1}^{m} \Box \Diamond \psi_i^{\mathrm{env}}$. So, let $l_1$ be the smallest value at which $\sigma_{l_1} \models \psi_1^{\mathrm{sys}}$ holds.

Now consider $\bar{\sigma}_{:k_1}$, the path to $\bar{\sigma}_{k_1}$. Let us look at the sequence $\sigma_{l_1:}$. If $\sigma_{l_1:} \not\models \Diamond \psi_2^{\mathrm{sys}}$, then the sequence $\bar{\sigma}_{:k_1} \sigma_{l_1:} \in$ $\mathrm{Plays}(f_G^{\bar{\varphi}})$. This is because $|[[\psi_1^{\mathrm{sys}}]]|=1$, $\bar{\sigma}_{k_1} = \sigma_{l_1}$. And by construction,

$$\bar{f}((m_{k_1}^{\sigma_{l_1}, \bar{f}}, 1), \sigma_{l_1}, \sigma_{l_1+1} \cap \mathrm{AP}_{\mathrm{env}}) = \\ f(m_{k_1}^{\bar{\sigma}, f_G^{\bar{\varphi}}}, \bar{\sigma}_{k_1}, \sigma_{l_1+1} \cap \mathrm{AP}_{\mathrm{env}}).$$

Therefore, $\bar{\sigma}_{:k_1} \sigma_{l_1:} \in \mathrm{Plays}(f_G^{\bar{\varphi}})$ and $(\bar{\sigma}_{:k_1} \sigma_{l_1:})_0 \in W_{\bar{\varphi}}$. This implies that

$$\bar{\sigma}_{:k_1} \sigma_{l_1:} \models \psi_1^{\mathrm{sys}} \wedge \Diamond(\psi_1^{\mathrm{sys}} \to \Diamond \psi_2^{\mathrm{sys}})$$

from the definition of $\bar{\varphi}$ and $m_{k_1}^{\bar{\sigma}, f_G^{\bar{\varphi}}}$ – leading to a conttradiction to our assumption $\sigma_{l_1:} \not\models \Diamond \psi_2^{\mathrm{sys}}$. Thus, there exists a finite $l_2 \geq l_1$ at which $\sigma_{l_2} \models \psi_2^{\mathrm{sys}}$. The inequality $l_2 \geq l_1$ can be made strict i.e $l_2 > l_1$ by identifying any $i, j$ for which $[[\psi_j^{\mathrm{sys}}]] = [[\psi_i^{\mathrm{sys}}]]$, and combining them into one progress condition. This means that the same state will not satisfy any two distinct progress conditions, hence $l_2 > l_1$ from the condition $\Diamond(\psi_1^{\mathrm{sys}} \to \Diamond \psi_2^{\mathrm{sys}})$.

Repeating the argument for any $i$, we get $\exists l_{i\oplus 1} > l_i$ such that $l_{i+1}$ is finite and $\sigma_{l_{i\oplus 1}} \models \psi_{i\oplus 1}^{\mathrm{sys}}$ with $\sigma_{l_i} \models \psi_i^{\mathrm{sys}}$. This way we showed that there exists a sequence of integers such that $l_1^1 < l_2^1 < \ldots l_n^1 < l_1^2 \ldots < l_j^k$ $\forall j \leq n, \forall k$ with $\sigma_{l_k^i} \models \psi_i^{\mathrm{sys}}$. Given any $j \leq n$ and $r \in \mathbb{N}$ we can find a $k$ such that $r < (k-1)n$, $\sigma_{l_j^k} \models \psi_j^{\mathrm{sys}}$. Therefore, $\sigma_{r:} \models \Diamond \psi_j^{\mathrm{sys}}$. This holds true for all $r$ and for all $j \leq n$, hence $\sigma \models \bigwedge_{i=1}^{n} \Box \Diamond \psi_i^{\mathrm{sys}}$. Therefore, $\bar{f}$ is winning against $\varphi$ and $\sigma_0 \in W_{\bar{\varphi}} \to \sigma_0 \in W_{\varphi}$. Therefore, $W_{\bar{\varphi}} \subseteq W_{\varphi}$. And from before $W_{\varphi} \subseteq W_{\bar{\varphi}}$, hence $W_{\bar{\varphi}} = W_{\varphi}$. ∎

## APPENDIX II
## PROOF OF LEMMA 3

*Proof:* We first show that for a game with $\bar{\varphi}$ as the winning condition, we can compute the winning strategy from solving $n+1$ reachability games. Then we use the result from Claim 2.

Consider a state $s$ that is winning for $\bar{\varphi}$. Let $\bar{f}_G^{\bar{\varphi}}$ be the winning strategy for $\bar{\varphi}$ from $s$. Consider $\sigma \in Plays(\bar{f}_G^{\bar{\varphi}})$, then $\sigma \models \psi_1^{\mathrm{sys}} \wedge \bigwedge_{j=1}^{n} \Diamond(\psi_j^{\mathrm{sys}} \to \Diamond \psi_{j\oplus 1}^{\mathrm{sys}})$ or $\sigma \models \bigvee_{i=1}^{m} \Diamond \Box \neg \psi_i^{\mathrm{env}}$. If all plays of $\bar{f}_G^{\bar{\varphi}}$ with the initial state as $s$ satisfy $\bigvee_{i=1}^{m} \Diamond \Box \neg \psi_i^{\mathrm{env}}$, then $\bar{f}_G^{\bar{\varphi}}$ is winning for $\varphi_0^{\mathrm{reach}}$ as well from $s$. Therefore, by solving the reachability game with $\varphi_0^{\mathrm{reach}}$ as the winning condition, we can obtain a strategy winning for $\varphi$.

Consider the case when $\exists \sigma \in \mathrm{Plays}(\bar{f}_G^{\bar{\varphi}})$ such that $\sigma \models \bigwedge_{i=1}^{m} \Box \Diamond \psi_i^{\mathrm{env}}$ and $\sigma_0 = s$. We observe that for this case, for each $i \in \{1, 2, \ldots, n\}$, the reachability game with condition $\varphi_i^{\mathrm{reach}}$ is winnable from $[[\psi_i^{\mathrm{sys}}]]$. To do this, first note

$$\sigma \models \psi_1^{\mathrm{sys}} \wedge \bigwedge_{j=1}^{n} \Diamond(\psi_j^{\mathrm{sys}} \to \Diamond \psi_{j\oplus 1}^{\mathrm{sys}})$$

since $\sigma \models \bigwedge_{i=1}^{m} \Box \Diamond \psi_i^{\mathrm{env}}$ and $\sigma \models \bar{\varphi}$. Define $r_i$ to be the smallest instance such that $\sigma_{r_i} \models \psi_i^{\mathrm{sys}}$ (we know that such an $r_i$ exists from the arguments in the proof of Claim 2).

Next, we prove that the strategy $\bar{f}_G^{\bar{\varphi}}$ with the initial memory value as $m_{r_i}^{\sigma, \bar{f}_G^{\bar{\varphi}}}$ is winning for game with condition $\varphi_i^{\text{reach}}$. To see this, consider any $\underline{\sigma} \in \text{Plays}(\bar{f}_G^{\bar{\varphi}})$ with $\underline{\sigma}_0 = q$ such that $q \models \psi_i^{\text{sys}}$ i.e execute according to the strategy starting at the state $q$ and memory $m_{r_i}^{\sigma, \bar{f}_G^{\bar{\varphi}}}$. Note that since $[[\psi_i^{\text{sys}}]]$ is a singleton, $q = \sigma_{r_i} = \underline{\sigma}_0$. And, at the state-memory value pair $(q, m_{r_i}^{\sigma, \bar{f}_G^{\bar{\varphi}}})$, $\bar{f}_G^{\bar{\varphi}}$ is well-defined since this state-memory value pair is reachable for $\bar{f}_G^{\bar{\varphi}}$ (recall we selected this state and memory value from a execution in $Plays(\bar{f}_G^{\bar{\varphi}})$ starting from the winning set for $\bar{\varphi}$).

Case 1: $\underline{\sigma} \models \bigwedge_{i=1}^{m} \Box\Diamond \psi_i^{\text{env}}$

For this case, $\sigma_{:r_i}\underline{\sigma} \in Plays(\bar{f}_G^{\bar{\varphi}})$ since $\forall k < r_i, (m_{k+1}^\sigma, \sigma_{k+1} \cap \text{AP}_{\text{sys}}) = \bar{f}_G^{\bar{\varphi}}(m_k^\sigma \sigma_k, \sigma_{k+1} \cap \text{AP}_{\text{env}})$ and since $\sigma_{r_i} = \underline{\sigma}_0$. We continue execution from $(\sigma_{r_i}, m_{r_i}^{\sigma, \bar{f}_G^{\bar{\varphi}}})$ in accordance with $\bar{f}_G^{\bar{\varphi}}$, so the entire sequence was generated in accordance with this strategy. Using the fact that this strategy is winning from $\sigma_0$ for $\bar{\varphi}$, and that the semantics of LTL imply

$$\underline{\sigma} \models \bigwedge_{i=1}^{m} \Box\Diamond\psi_i^{\text{env}} \to \sigma_{:r_i}\underline{\sigma} \models \bigwedge_{i=1}^{m} \Box\Diamond\psi_i^{\text{env}},$$

we arrive at the conclusion that

$$\sigma_{:r_i}\underline{\sigma} \models \psi_1^{\text{sys}} \wedge \bigwedge_{j=1}^{n} \Diamond(\psi_j^{\text{sys}} \to \Diamond\psi_{j\oplus 1}^{\text{sys}}).$$

Therefore, $\sigma_{:r_i}\underline{\sigma} \models (\psi_i^{\text{sys}} \wedge \Diamond\psi_{i\oplus 1}^{\text{sys}})$ and $r_i$ was the smallest instance at which $\psi_i^{\text{sys}}$ holds. It follows that $\underline{\sigma} \models (\psi_i^{\text{sys}} \wedge \Diamond\psi_{i\oplus 1}^{\text{sys}})$. Therefore, $\underline{\sigma} \models \varphi_i^{\text{reach}}$.

Case 2: $\underline{\sigma} \models \neg \bigwedge_{i=1}^{m} \Box\Diamond\psi_i^{\text{env}}$

$\underline{\sigma} \models \neg \bigwedge_{i=1}^{m} \Box\Diamond\psi_i^{\text{env}} \to \underline{\sigma} \models \varphi_i^{\text{reach}}$.

This implies that all plays of $f_G^{\bar{\varphi}}$ starting with $s \models \psi_i^{\text{sys}}$ and initial memory value $m_{r_i}^{\sigma, \bar{f}_G^{\bar{\varphi}}}$ are winning against $\varphi_i^{\text{reach}}$. Hence, $f_G^{\bar{\varphi}}$ is winning against $\varphi_i^{\text{reach}}$ for the state $s \models \psi_i^{\text{sys}}$. For the case with $s \models \theta$, by the definition of $\bar{\varphi}$, the set of states $[[\theta]]$ is winning for $\varphi_n^{\text{reach}}$.

Now, we have shown that for the case when the reachability game $\varphi_0^{\text{reach}}$ is not winnable, if $\bar{\varphi}$ is winnable, we can find a winning strategy for each of the $\varphi_j^{\text{reach}}$ games with their respective initial conditions as described in Section III. Let the winning strategy for each such reachability game be $f_G^{\text{reach}_j} : M^j \times \Sigma \times \mathcal{P}(\text{AP}_{\text{env}}) \to M^j \times \mathcal{P}(\text{AP}_{\text{sys}})$ with $m_0^i$ as the initial memory. Without loss of generality, assume that the for any $i, j$ with $i \neq j$, $M^i \cap M^j = \emptyset$. The earlier segment of the proof was to show the existence of these strategies when $\bar{\varphi}$ is winnable.

Now we show these can be combined to form a winning strategy $f_G^{\bar{\varphi}}$ winning against $\bar{\varphi}$. First, consider the strategy $f^{\text{reach}_n}$. Replace all the memory values corresponding to reachable $(w, s)$ for $f^{\text{reach}_n}$ where $s \models \psi_1^{\text{sys}}$ with $m_0^1$. Note that $s \models \psi_1^{\text{sys}}$ corresponds to a valid initial state for the game with condition $\varphi_1^{\text{reach}}$. Let this modified strategy be $\bar{f}^{\text{reach}_n}$. Effectively, we have patched $f^{\text{reach}_n}$ with $f^{\text{reach}_1}$ so that after reaching a state that satisfies $\psi_1^{\text{sys}}$ it switches from $f^{\text{reach}_n}$ to $f^{\text{reach}_1}$. Call the new resulting strategy $f_{1,2}^*$ where $f_{1,2}^*$ is defined as:

$$f_{1,2}^*(w', y) = \begin{cases} \bar{f}^{\text{reach}_n}(w, s, x) & \text{if } w \in M^n, \\ f^{\text{reach}_1}(w, s, x) & \text{if } w \in M^1. \end{cases}$$

Consider $\sigma$ in $\text{Plays}(f_{1,2}^*)$ such that $\sigma \models \theta \wedge \bigwedge_{i=1}^{m} \Box\Diamond\psi_i^{\text{env}}$ (the other case is trivial). Initially we start at memory $m_0^n$ and a state $\sigma_0 : \sigma_0 \models \theta$ and continue execution along strategy $f^{\text{reach}_n}$ till we reach $\psi_1^{\text{sys}}$ in a finite number of steps–this is guaranteed by the definition of $\varphi_n^{\text{reach}}$. Subsequently, execution is continued along $f^{\text{reach}_1}$ till we reach a state that satisfies $\psi_2^{\text{sys}}$ in a finite number of steps. This is because if $\sigma \models \bigwedge_{i=1}^{m} \Box\Diamond\psi_i^{\text{env}}$, then $\sigma \models \Diamond\psi_1^{\text{sys}} \wedge \Diamond(\psi_1^{\text{sys}} \to \Diamond\psi_2^{\text{sys}})$. Otherwise from the definition of $f_1^{\text{reach}}$ and $\varphi_1^{\text{reach}}$, we end up with a contradiction as before. Similarly, we extend $f_{1,2}^*$ to replace the memory corresponding to the reachable $(w, s)$ for $f_1^{\text{reach}}$ where $s \models \psi_2^{\text{sys}}$ with $m_0^2$. Define the resulting strategy $f_{1,2,3}^*$ as:

$$f_{1,2,3}^*(w', y) = \begin{cases} \bar{f}^{\text{reach}_n}(w, s, x) & \text{if } w \in M^n, \\ \bar{f}^{\text{reach}_1}(w, s, x) & \text{if } w \in M^1, \\ f^{\text{reach}_2}(w, s, x) & \text{if } w \in M^2. \end{cases}$$

As before, we can show that the plays of this strategy are winning against $\Diamond\psi_1^{\text{reach}} \wedge \Diamond(\psi_1^{\text{reach}} \to \Diamond\psi_2^{\text{reach}}) \wedge \Diamond(\psi_2^{\text{reach}} \to \Diamond\psi_3^{\text{reach}})$. Continue the procedure to obtain $f_{1,2,\ldots,n,1}^*$. By construction, this stategy is winning against $\bar{\varphi}$.

We argued that for a state $s \models \theta$ that is winning for $\bar{\varphi}$, either the reachability game with condition $\varphi_0^{reach}$ is winnable or the reachability games with condition $\varphi_i^{reach}$ with $i \in \{1, 2, \ldots, n\}$ are winnable. And for both cases, we provided a construction for a strategy winning against $\bar{\varphi}$ from the strategies winning for the reachability games. This implies that for a state, $\bar{\varphi}$ is winnable if and only if the game with $\varphi_0^{reach}$ is winnable or the games with $\varphi_i^{reach}$ are winnable. Therefore, from solving the reachability games we can infer if a state is winnable or not and also construct a winning strategy for $\bar{\varphi}$ if it is winnable.

In the proof of Claim 2, we demonstrated an approach to construct a strategy that is winning against $\varphi$ using the strategy winning against $\bar{\varphi}$. We also showed that the winning states for the conditions $\varphi$ and $\bar{\varphi}$ are the same. Hence, the GR(1) game can be solved by solving $n+1$ reachability games separately. ∎